\documentclass[conference]{IEEEtran}
\IEEEoverridecommandlockouts
\usepackage{cite}
\usepackage{amsmath,amssymb,amsfonts}
\usepackage{algorithmic}
\usepackage{graphicx}
\usepackage{textcomp}
\usepackage{xcolor}
\usepackage{bbm}
\usepackage{braket}
\usepackage{bm}
\usepackage{hyperref}
\usepackage[capitalize]{cleveref}
\usepackage{comment}

\newcommand{\ms}{\mathrm{ms}}

\def\BibTeX{{\rm B\kern-.05em{\sc i\kern-.025em b}\kern-.08em
    T\kern-.1667em\lower.7ex\hbox{E}\kern-.125emX}}

\newcommand{\quotes}[1]{``#1''}

\begin{document}

\title{
Q-Profile: Profiling Tool for Quantum Control Stacks
applied to the Quantum Approximate Optimization Algorithm
}

\author{\IEEEauthorblockN{Koen J. Mesman}
\IEEEauthorblockA{
\textit{Qblox}\\
Delft, Netherlands \\
koenmesman@gmail.com}
\and
\IEEEauthorblockN{Francesco Battistel}
\IEEEauthorblockA{
\textit{Qblox}\\
Delft, Netherlands \\
fbattistel@qblox.com}
\and
\IEEEauthorblockN{Edgar Reehuis}
\IEEEauthorblockA{
\textit{Qblox}\\
Delft, Netherlands \\
ereehuis@qblox.com}
\and
\IEEEauthorblockN{Damaz de Jong}
\IEEEauthorblockA{
\textit{Qblox}\\
Delft, Netherlands \\
ddejong@qblox.com}
\and
\IEEEauthorblockN{Marijn J. Tiggelman}
\IEEEauthorblockA{
\textit{Qblox}\\
Delft, Netherlands \\
mtiggelman@qblox.com}
\and
\IEEEauthorblockN{Jordy Gloudemans}
\IEEEauthorblockA{
\textit{Qblox}\\
Delft, Netherlands \\
jgloudemans@qblox.com}
\and
\IEEEauthorblockN{Jules C. van Oven}
\IEEEauthorblockA{
\textit{Qblox}\\
Delft, Netherlands \\
jules@qblox.com}
\and
\IEEEauthorblockN{Cornelis C.~Bultink}
\IEEEauthorblockA{
\textit{Qblox}\\
Delft, Netherlands \\
niels@qblox.com}
}

\maketitle

\begin{abstract}
Scaling up the number of qubits and speeding up the execution of quantum algorithms are important steps towards reaching quantum advantage.
This poses heavy demands particularly on the control stack, as pulses need to be distributed to an increasing number of control channels and variational algorithms require rapid interleaving of quantum and classical computation.
Assessing the bottlenecks in the control stack is therefore key to making it ready for reaching quantum advantage.
However, existing benchmark suites suffer from lack of detail due to indirect access to the control hardware.
In this work, we present Q-Profile, a tool to profile quantum control stacks that circumvents these issues by utilizing a direct connection from the host CPU to the control stack, providing fine accuracy in measuring the runtime and allowing to identify performance bottlenecks.
We demonstrate the use of our tool by benchmarking the Quantum Approximate Optimization Algorithm (QAOA) on a Qblox Cluster for a virtual 4 to 14-qubit transmon system.
Our results identify the major execution bottlenecks in the passive qubit reset and communication overhead.
We estimate a 1.40x~speedup with respect to the total runtime by using an active qubit reset, instead of passive reset,
and demonstrate a further speedup of 1.37x by parallel initialization of the control modules.
The presented method of profiling is applicable to other control-stack providers, as well as to other benchmarks, while still providing detailed information beyond a single metric.
By extension, this tool will enable identifying and eliminating bottlenecks for future quantum acceleration. 
The profiling tool is included in the open-source Quantify quantum control software, which allows support for multiple back-ends.

\end{abstract}


\section{Introduction}
\label{sec:intro}

In recent years, quantum computing has been growing at a fast rate towards executing a computational task faster than any classical computer~\cite{Arute19}.
The next milestone, now sought after, is achieving fast computations for practical utility, dubbed quantum advantage.
In order to achieve this, the number of controlled qubits has to scale up and the gate error rates have to decrease, each by multiple orders of magnitude. 
This places heavy demands on the quantum control stack (both in terms of hardware and software), where programs and pulses need to be distributed to an increasing number of control channels, and where hybrid optimization algorithms~\cite{farhi2014quantum,fedorov2021vqe} and gate-tuning protocols~\cite{Klimov20} require rapid interleaving of quantum-circuit execution, analysis and classical computation.
Furthermore, previous results~\cite{mesman2021qpack} indicate the importance of accelerating variational quantum algorithms, since worst-case runtimes of over 15~hours have been measured for minimal (5~qubit) example problems.
Assessing and understanding the bottlenecks in the control stack is therefore key to making it ready for reaching quantum advantage. 
While a number of benchmark suites have been developed in the last few years~\cite{mesman2021qpack, fermionic2020, atos2021, 2021QEDC}, they suffer from inaccuracies due to indirect access and connection to the control hardware.

In this work we present Q-Profile, a tool to benchmark quantum control stacks that circumvents long-distance high-latency connections by utilizing a direct connection from the host CPU to the quantum control stack. 
Having access to the control hardware runtime of the scheduled instructions, this tool provides fine detail and accuracy in measuring the quantum runtime. 
The profiling method is aimed at practical benefit, allowing control-stack developers to identify performance bottlenecks while executing near-term quantum algorithms on their systems.

To showcase Q-Profile, the quantum approximate optimization algorithm (QAOA)~\cite{farhi2014quantum} implementation from the QPack benchmark suite~\cite{mesman2021qpack}, a quantum application benchmark suite, is used. 
Using this implementation, instead of providing a benchmark score, we present a means of profiling the quantum control hardware, demonstrated on a Qblox Cluster for a 4 to 14 qubit virtual transmon system.
This means that the Cluster operates as if connected to actual transmon qubits, while only the binary measurement outcomes used by the QAOA~optimizer on the host CPU are simulated.
The QPack suite was applied for profiling IBM~hardware in earlier work~\cite{mesman2021qpack}, however, insufficient access to the underlying hardware gave unsatisfactory results. 
Instead, by having direct access to the control hardware, cloud communication and queue times are non-existent, greatly speeding up the processes of profiling.
Our findings identify the major execution bottlenecks in the modality of qubit reset (passive versus active) and communication overhead. 
For each of these bottlenecks, we propose and implement mitigation strategies, achieving a 1.40x~speedup for active reset compared to the baseline measurements, and a further 1.37x~speedup by parallel module initialization.
This method of profiling is applicable to other control-stack providers, as well as to other benchmarks. 
The profiling tool is included in the open-source Quantify~\cite{quantify} quantum control software, which allows support for multiple back-ends.
Additionally, Q-Profile can be used to measure CLOPS (Circuit-Layer Operations Per Seconds)~\cite{CLOPS} by running quantum-volume-like circuits instead of QAOA circuits, while still providing detailed information beyond a single metric.
By profiling the most demanding computation steps, developers can efficiently locate and mitigate performance issues in both software and hardware. 
By extension, this tool will enable identifying and eliminating future bottlenecks for quantum acceleration in general.

We give a short introduction to the quantum hardware structure~(\ref{sec:hardware}), a description of the quantum algorithm we use to evaluate the hardware~(\ref{sec:algorithms}) and the used metrics~(\ref{sec:benchmark_overall}). 
The results of the performance measurements are presented~(\ref{sec:results}) and discussed~(\ref{sec:discussion}), including potential strategies to increase performance. In~\cref{sec:conclusion} we summarize our findings.

\section{Quantum control hardware structure}
\label{sec:hardware}

\begin{figure}
    \centering
    \includegraphics[width=\linewidth]{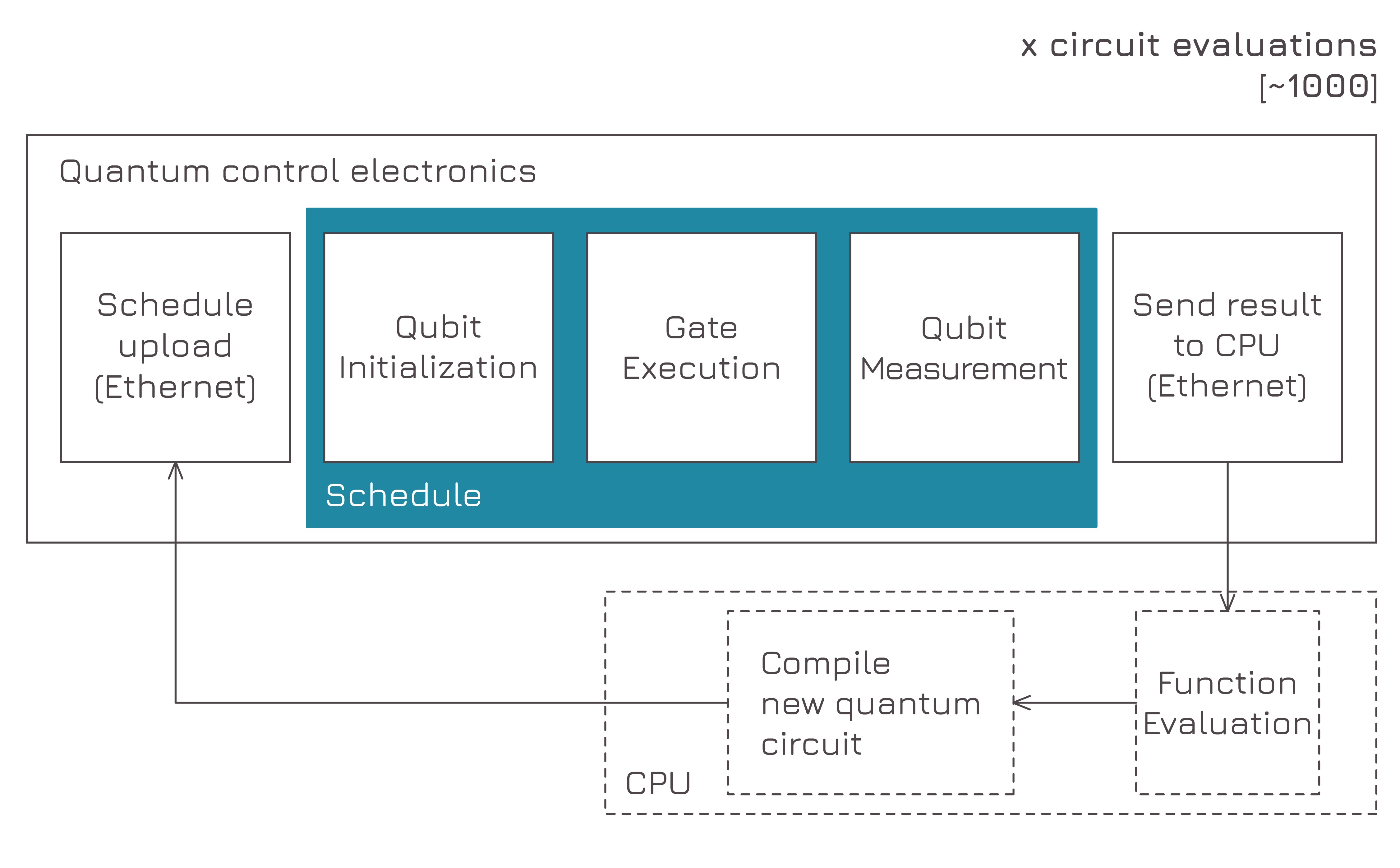}
    \caption{General layout of quantum computations.
        A host CPU compiles a quantum circuit into an executable schedule of operations.
        This is uploaded (e.g.~via Ethernet) to the quantum control hardware, which executes the schedule (initialization, gates, measurements) by sending appropriate pulses to the qubits for each of these operations.
        The circuit is usually repeated many times to collect statistics and the results are eventually sent back to the host CPU for post-processing. 
    }
    \label{fig:hw_layout}
\end{figure}

Quantum computers can vary significantly between implementations. 
Especially for quantum-computing chips, no consensus has been reached on which technology will become the standard~\cite{material_2021}.
However, in terms of the quantum-computing \emph{control}-hardware structure, many similarities can be found.
The general structure for executing quantum computations is given in~\cref{fig:hw_layout}.
A host classical CPU is required for general-purpose computations, pre- and post-processing.
The host CPU compiles a high-level quantum algorithm to executable quantum circuit instructions.
These are transferred to the quantum control hardware, e.g.~via Ethernet.
The control hardware executes gates and measurements in different manners, depending on the quantum-chip technology.
The control hardware generally stores, reads and interprets the received instructions in order to generate the respective pulse sequences.
Furthermore, the measurement results need to be stored, read and optionally processed before sending the results back to the host CPU. 
Each execution of a quantum circuit with a final measurement is referred to as a shot. 
In order to collect accurate statistics, thousands or more shots are generally executed for each instance of a quantum circuit.
The measurement results are either transferred back at the end of each shot or are temporarily saved locally, allowing to reduce latency.

\section{Hybrid quantum algorithms} 
\label{sec:algorithms}

In the current era of quantum computing, the  so-called NISQ~era (Noisy Intermediate-Scale Quantum), qubits are limited in number and quality, limiting the scope of problems that can be solved on current quantum computers.
Algorithms such as the well-known Shor's factorization or Grover's search algorithms require a large number of qubits to calculate practical use cases and need even more qubits to mitigate errors in the form of quantum error correction (20~million qubits are expected to be required to break standard RSA~cryptography~\cite{Gidney2021howtofactorbit}). 

In recent trends, hybrid classical/quantum algorithms are preferred~\cite{cerezo2021variational}. 
These algorithms are parametrized quantum algorithms which interact with classical optimizers to update and optimize their parameters.
Examples of these algorithms are QAOA~(Quantum Approximate Optimization Algorithm)~\cite{farhi2014quantum} and VQE~(Variational Quantum Eigensolver)~\cite{peruzzo2014variational}. 
One of the qualities that makes these algorithms appealing for NISQ~computing is that the number of qubits required to achieve quantum advantage might be of the order of a few hundreds to thousands of qubits~\cite{guerreschi2019qaoa}. 
While the current state of quantum-computing chips is not at this stage, it is expected to be reached in the near future~\cite{roadmap}.

While these algorithms seem promising, practical difficulties surface as well. 
Most prominent is the difficulty of classically optimizing the parameters for QAOA, as this optimization has been shown to be NP-hard~\cite{bittel2021training}.
Furthermore, since this algorithm includes both classical and quantum computing, the pipeline within the quantum control stack must be assessed to ensure efficient calculations.
\Cref{sec:results} will go into further detail regarding this assessment, and identify bottlenecks in the quantum control stack.

\begin{figure}
    \centering
    \includegraphics[width=\linewidth]{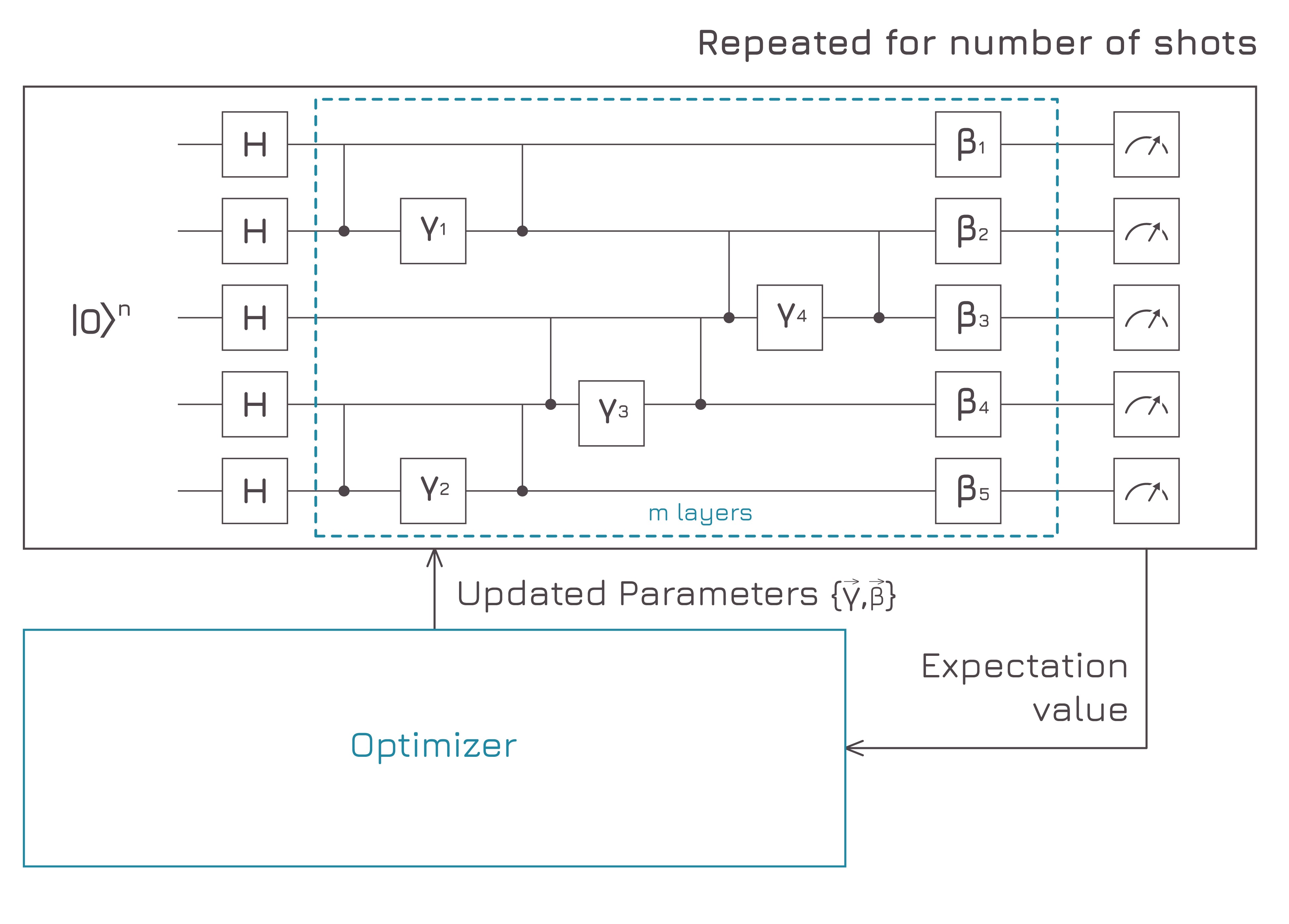}
    \caption{
        Graphical representation of the layout of QAOA.
        A classical computer updates the parameters~$\vec{\beta}$ and $\vec{\gamma}$ of QAOA.
        After each iteration, the measurement statistics is used to evaluate the cost function of the problem instance, and the QAOA~parameters are consequently updated by a classical optimizer.
        }
    \label{fig:qaoa}
\end{figure}

\subsection{Quantum Approximate Optimization Algorithm (QAOA)}
\label{sec:qaoa}

QAOA is a variational quantum algorithm proposed in Ref.~\cite{farhi2014quantum} intended to solve combinatorial optimization problems. 
QAOA is a layered algorithm. 
The layers consist of a \quotes{cost} unitary followed by a \quotes{mixer} unitary, each of which is implemented as a layer of parameterized gates.
The general idea is that the cost unitary encodes the problem instance to be solved, while the mixer Hamiltonian creates superpositions of the basis states in order to subsequently find the state which produces the best solution to the problem.
The optimization is a bound constrained optimization since the parameters are rotation angles bounded within~$[0, 2 \pi]$.
A representation of the general structure of the algorithm is shown in~\cref{fig:qaoa}. 
Note that only the single-qubit gates are parametrized, whereas the two-qubit gates are static.
The implementation used in this work uses 2~layers, as deemed sufficient for small problem sizes~\cite{qaoa_layers}. 
Furthermore, in a noise-free scenario, increasing the number of layers increases the accuracy~\cite{farhi2014quantum}, however this may not be the case on noisy quantum computers.
We note that while here the number of layers stays constant for all problem sizes, the circuit depth per layer can increase as a function of the problem size.
In general, a constraint can be set between 2~qubits, which translates to a set of two-qubit gates.
In the worst-case scenario, all qubits share a constraint with all other qubits, which results in $\frac{n(n-1)}{2}$ constraints, where $n$ is the number of qubits.
In the used benchmark set, the ratio of constraints to qubits remains constant (4~constraints per qubit) and therefore the growth is linear.

\section{Control-hardware benchmark}
\label{sec:benchmark_overall}

\subsection{Benchmark metrics}
\label{sec:benchmark}

To benchmark quantum control hardware, several metrics must be evaluated, including the runtime performance and the result accuracy, as outlined and implemented in earlier work~\cite{mesman2021qpack}.
In this paper we focus on the runtime performance. 
The result accuracy is relevant for users of our tool that have access to a quantum chip in an experimental setup.
However, in this work a subroutine simulates the noiseless quantum circuit (like the one in~\cref{fig:qaoa}) to make the benchmarking and profiling reproducible and independent of the availability of a cryostat and quantum device calibration.
As such, while our tool can evaluate the result accuracy as well, we do not discuss it in the rest of this work.

The main focus of the benchmark tool that we introduce here is a fine-grained resolution of the runtime for the various computing steps.
The measured steps are:

\begin{itemize}
    \item Compile: quantum-circuit compilation to Q1ASM~instructions by the control software on the host CPU (Qblox uses the open-source multi-vendor Quantify software~\cite{quantify}).
    \item Schedule (execution): execution of the circuit (gates, measurements) and qubit reset as pulses on a (virtual) quantum chip.
    \item Communication: communication from the host CPU to the quantum control hardware and back, including possibly multiple rounds to set up the modules and send the compiled instructions. The communication category is further subdivided into:
    \begin{itemize}
        \item Stop: command to stop the schedule.
        Used as a safety precaution both to make sure the modules are ready before receiving new schedules, and after receiving the measured data.
        \item Prepare: sending the compiled waveform schedule over an Ethernet connection to the control and readout modules, which then make the sequencers ready.
        \item Start: command to start the waveform schedule.
        \item Wait-done: host CPU waiting and checking until the schedule has finished, net of the schedule-execution time 
        (in the profiling measurements we performed, this runtime includes the schedule-execution runtime, however the latter is subtracted to isolate the wait-done overhead).
        \item Retrieve acquisition: receiving the measured results from the readout modules to the host CPU.
    \end{itemize}
    \item Total: the total runtime of the hybrid quantum/classical algorithm.
\end{itemize}

The current implementation gives several advantages.
First, as there is a direct connection from the host CPU to the quantum control hardware, there are no factors such as API~calls or long-distance high-latency connections.
Secondly, fine control in measuring the quantum runtime is available.
It is indeed possible to directly access the pipeline of all scheduled instructions, allowing for an exact measurement of the runtime.
Furthermore, one can also go into detail on different sub-processes of the major computational steps.

\subsection{Methods}
\label{sec:methods}

\begin{figure}
    \centering
    \includegraphics[width=\linewidth]{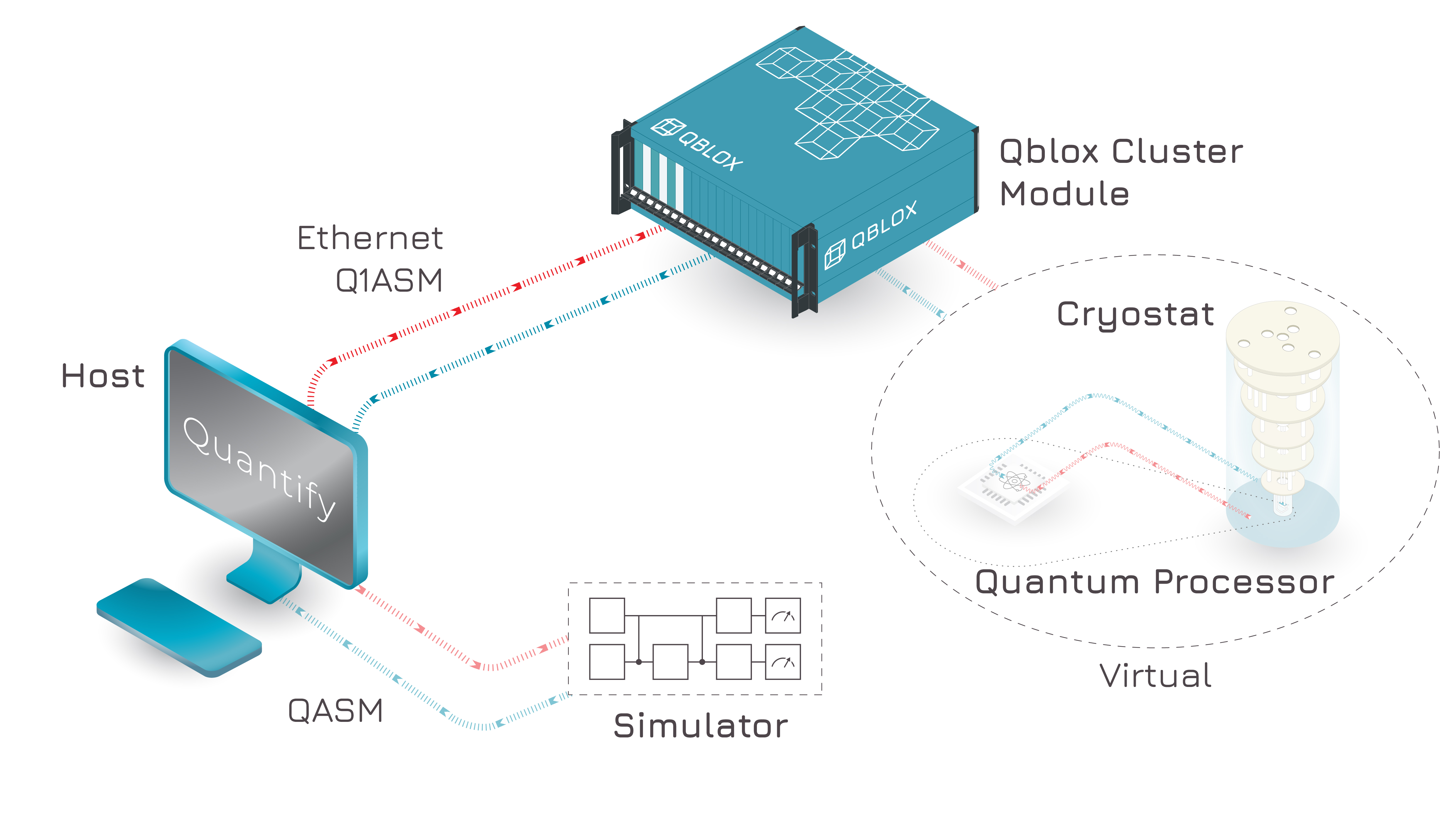}
    \caption{Used initialization for benchmarking the control hardware.
    A host CPU communicates Q1ASM~instructions to a Qblox Cluster via Ethernet, which performs its operations as if connected to actual transmon qubits.
    The random measurement results reported by the Cluster for a virtual chip are substituted in the host CPU with the measurement results obtained using QASM to simulate the QAOA~circuits like the one in~\cref{fig:qaoa}.}
    \label{fig:initialisation}
\end{figure}

The initialization used in this work for benchmarking and profiling uses a host CPU (Intel Core i5 10th Gen i5-10210U, 8GB RAM, Nvidia GeForce MX250) connected to a Qblox Cluster through a 1~Gbps Ethernet connection (see~\cref{fig:initialisation}).
For 14~qubits, in this work the Cluster uses three quantum control modules and three readout modules to generate the pulses for single-qubit gates and measurements.
As two-qubit gates are static, and thus their schedule is always the same, we do not actually generate the corresponding pulses in separate control modules (but we account for their gate time in the reported schedule times).
Each module employs a time-deterministic and Turing-complete processor, named Q1~processor, for executing the control and readout tasks.
Q1~processors are programmed with the Q1ASM~assembly code that is compiled by Quantify on the host CPU. 
In order to faithfully reproduce experimental conditions and estimate the runtime,
the Cluster performs its operations as if it were connected to actual transmon qubits.
This means that the Cluster generates pulses and measures the signal at its input ports. 
As these are disconnected in this scenario, the Cluster effectively measures noise, communicating random measurement results in the retrieve-acquisition step.
These measurements are returned to the host CPU, but are immediately discarded and substituted with the measurements from the simulated circuit, which are then used by the optimizer for the parameter updates.

The circuit simulation is implemented through the Qiskit~Aer simulator~\cite{quasmAer}, but for future usage any QASM-supporting simulator can be integrated.
Note that feedback on the measurement results is required for running a hybrid quantum algorithm such as QAOA, hence the simulation is required to run the benchmark (of course, one can use an actual quantum device, if available).
The considered device parameters for gates and operations are reported in~\cref{tab:gates} and are inspired by Ref.~\cite{Krinner21}.
Furthermore, we consider an all-to-all connectivity for simplicity of compilation (see~\cref{sec:discussion} for an estimate of the required overhead for using SWAP gates).

\begin{table}
    \centering
    \caption{Considered execution times for quantum operations.}
    \begin{tabular}{|c|c|c|c|c|}
    \hline
        Passive reset & Active reset & 1Q gate & 2Q gate & Measurement \\
        200~$\mu$s & 1~$\mu$s &  40~ns & 100~ns & 500~ns\\
        \hline
    \end{tabular}
    \label{tab:gates}
\end{table}

The QAOA~algorithm is implemented using the QPack suite~\cite{mesman2021qpack}. 
Specifically, we consider QAOA for the Max-Cut problem for 4 to 14~qubits. 
For each qubit number, we consider a single problem instance and we run the benchmark for 40~times~starting from the same initial parameters.
As the (simulated) measurement outcomes are probabilistic, the number of optimization iterations to reach convergence varies each time.
To take this into account, the time results in~\cref{sec:results} are averaged over the number of iterations and runs.
The classical optimizer used is the SHGO~optimizer (Simplicial Homology Global Optimization)~\cite{shgo} included in the \emph{scipy.optimize} library~\cite{scipy}.
We note that it takes 0.47~ms per iteration for 4~qubits.
We use 1000~shots for each iteration, as this order of shots has proven to give sufficiently accurate optimization results for the examined small problem sizes~\cite{mesman2021qpack}. 
The QAOA~schedule uses single- and two-qubit gates for a maximum total of 64~gates, with a final measurement, resulting in a total maximum circuit time of 14.8~$\mu$s.
However, the compiled schedules will generally have lower execution times, as gates for which the parameters are set to~0 by the optimizer will be cancelled or merged by the compiler.

\section{Results}
 \label{sec:results}

 \subsection{Benchmark on 4~qubits and addressed bottlenecks}
 \label{sec:results_4Q}

\begin{figure*}
    \centering
    \includegraphics[trim=20 20 45 35, clip,width=\linewidth]{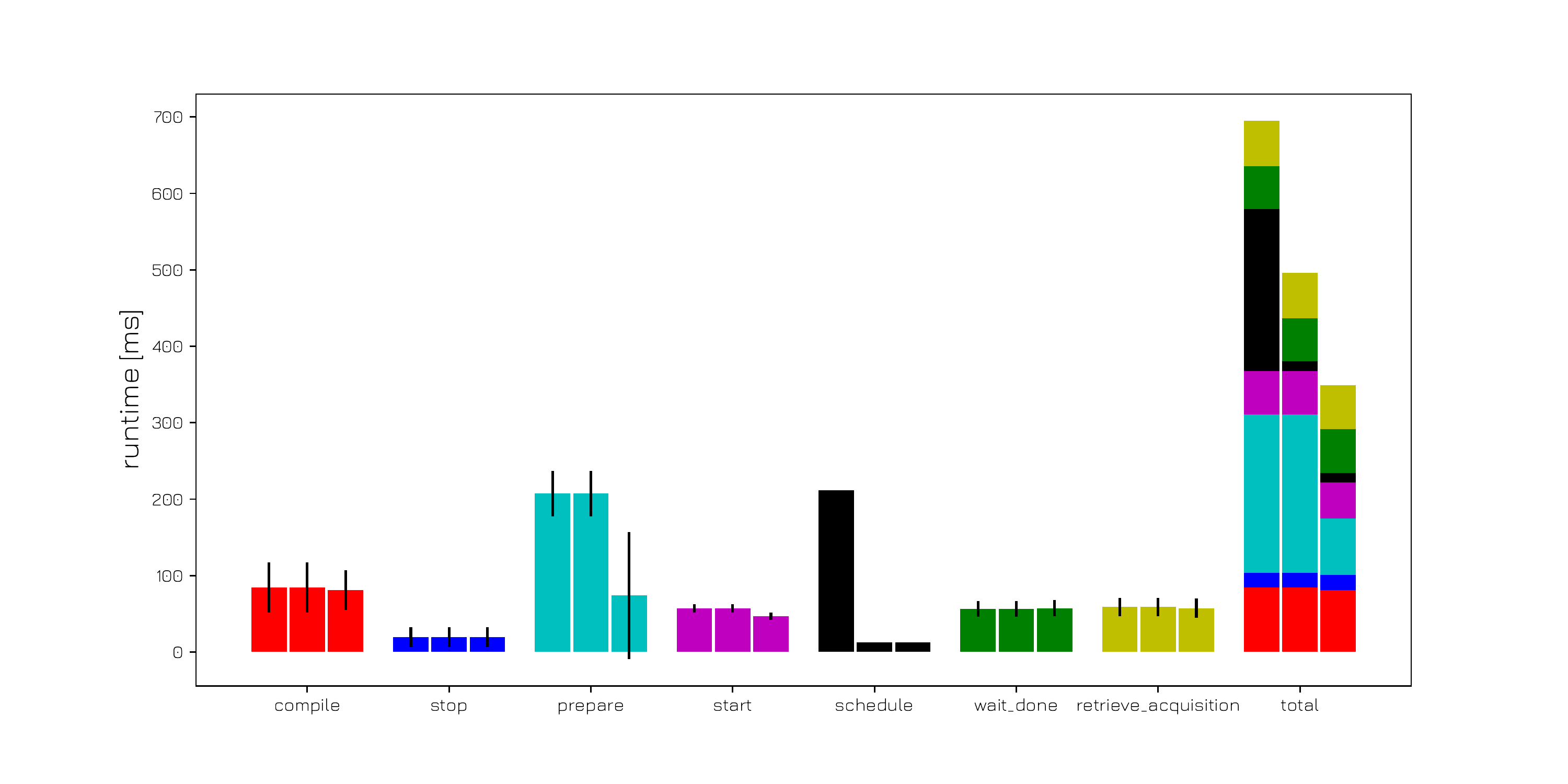}
    \caption{
    Average runtimes per optimization iteration for QAOA on 4~qubits, each measured for a total of 1000~shots (see~\cref{tab:result_table} for the exact values).
    Specifically, the results are averaged over the optimization iterations and over 40 runs of the benchmark.
    The error bars represent the standard deviation (root-mean-square deviation).
    The figure shows the runtimes for baseline (left column), using an active reset (middle) and for both an active reset combined with the implemented parallel module preparation.
    Note that the parallel module preparation is independent of using a passive or active reset.
    }
    \label{fig:bench_res}
\end{figure*}

\begin{table}
    \centering
    \caption{Average runtimes per optimization iteration for QAOA on 4~qubits for 1000 shots, shown in~\cref{fig:bench_res}.
   }
    \begin{tabular}{|c|c|}
        \hline
        \textbf{Computation step} & \textbf{Measured runtime [ms]}\\
        \hline
        \hline
        Circuit execution &  11.7 \\
        \hline
        Reset (passive) & 200  \\
        Reset (active) & 1  \\
        \hline
        Schedule execution (passive) &  211.7 \\ 
        Schedule execution (active) &  12.7 \\ 
        \hline
        \hline
        Start  & 57.11 \\
        \hline
        Prepare (baseline)  & 207.2  \\
        Prepare (parallel)  & 73.9  \\
        \hline
        Wait done  & 56.3 \\
        \hline
        Retrieve acquisition  & 58.9   \\
        \hline
        Stop  & 19.5 \\
        \hline
        Communication total (baseline) & 388.1 \\
        Communication total (parallel) & 254.8 \\
        \hline
        \hline
        Compilation & 84.2 \\
        \hline
        Optimizer & 0.47 \\
        \hline
        \hline
        Total (baseline) & 694.8\\
        Total (active reset) &  495.7 \\
        Total (active reset + parallel)  & 348.8 \\
        \hline
    \end{tabular}
    \label{tab:result_table}
\end{table}

The circuit execution (i.e.~gates, measurements, excluding qubit reset) forms the essential core of a quantum algorithm or experiment in general.
However, the profiling results in~\cref{fig:bench_res} (see also~\cref{tab:result_table}) show that the main part of the computation is \emph{not} spent on the circuit execution.
Indeed, one can compute that the baseline total runtime is 59.4~times the circuit-execution runtime (694.8~ms versus 11.7~ms).
Hence, reducing this huge overhead will have a significant impact for real-world applications.
The most significant steps in the computation for baseline are, in descending order, the qubit reset if passive (as it is often the case), the communication between the host CPU and the quantum control hardware (specifically the module preparation step), and the schedule compilation.
We note that the runtime standard deviation per run can be significant -- for example, for the longest run we measured 24.49~s for a total of 34~iterations and for the shortest 5.62~s for 8~iterations.
Especially for large or noisy systems, large amounts of iterations are required for the classical optimizer. 
As such, it is most relevant to consider the average runtimes per iteration.

We first discuss how to mitigate the bottleneck due to the passive qubit reset.
This can be done by implementing an active qubit reset instead~\cite{active_reset}.
We assume that the active reset will take approximately the qubit measurement time, plus the time for a corrective gate, for which we use a conservative estimate of 1~$\mu$s overall.
At the time of writing, feedback cycles with the Cluster are shortened to 320~ns, allowing for 500-700~ns reset cycles (depending on the time of flight and readout pulse duration).
This increases the schedule execution speed by an estimated
\begin{equation}
    \frac{211.7}{211.7 - 199} = 16.67~\text{times},
\end{equation}
leading to an overall speedup of
\begin{equation}
    \frac{694.8}{694.8 - 199} = 1.40~\text{times}.
\end{equation}
The comparison to the baseline is given visually in~\cref{fig:bench_res} active reset (middle columns).

The second execution bottleneck that emerged is the communication overhead~(388.1~ms). 
This is much more than the measured, bare latency of 2.5-3~ms over a short Ethernet cable.
We observed that this overhead is mostly due to the initialization (\quotes{preparation}) of the control and readout modules in the Cluster.
The control software (here Quantify) prepares these modules sequentially.
We implemented a parallelization of this preparation step through multi-threading.
In particular, instead of loading the compiled waveforms sequentially per module, the waveforms are loaded in parallel.

If parallelization would be complete, the runtime for parallelized preparation would be
\begin{equation}
   388.1~\ms - \frac{(388.1~\ms - 2.5~\ms) \times 4}{5} = 79.6~\ms.
\end{equation}
However, this maximum speedup is not achieved and is attributed to non-parallelizable overhead.

The actual improvement is visualized in~\cref{fig:bench_res} active reset +~parallel module initialization (right columns).
The measured speedup with this implementation is then:
\begin{equation}
    \frac{694.8}{694.8 - (388.1-254.8)} = 1.24~\text{times} ,
\end{equation}
whereas if this were implemented on top of an active qubit reset, where the overhead bottleneck is more apparent, the speedup is
\begin{equation}
    \frac{694.8-199}{694.8 - 199 - (388.1-254.8)} = 1.37~\text{times}.
\end{equation}

Note in~\cref{fig:bench_res} that the parallel prepare step displays a relatively large standard deviation. 
We suspect it to be caused by conflicts during communication, as this can occur when employing multi-threading.

\subsection{Scaling from 4 to 14~qubits}
\label{sec:results_14Q}

\begin{figure}
    \centering
    \includegraphics[trim=15 15 8 30, clip,width=\linewidth]{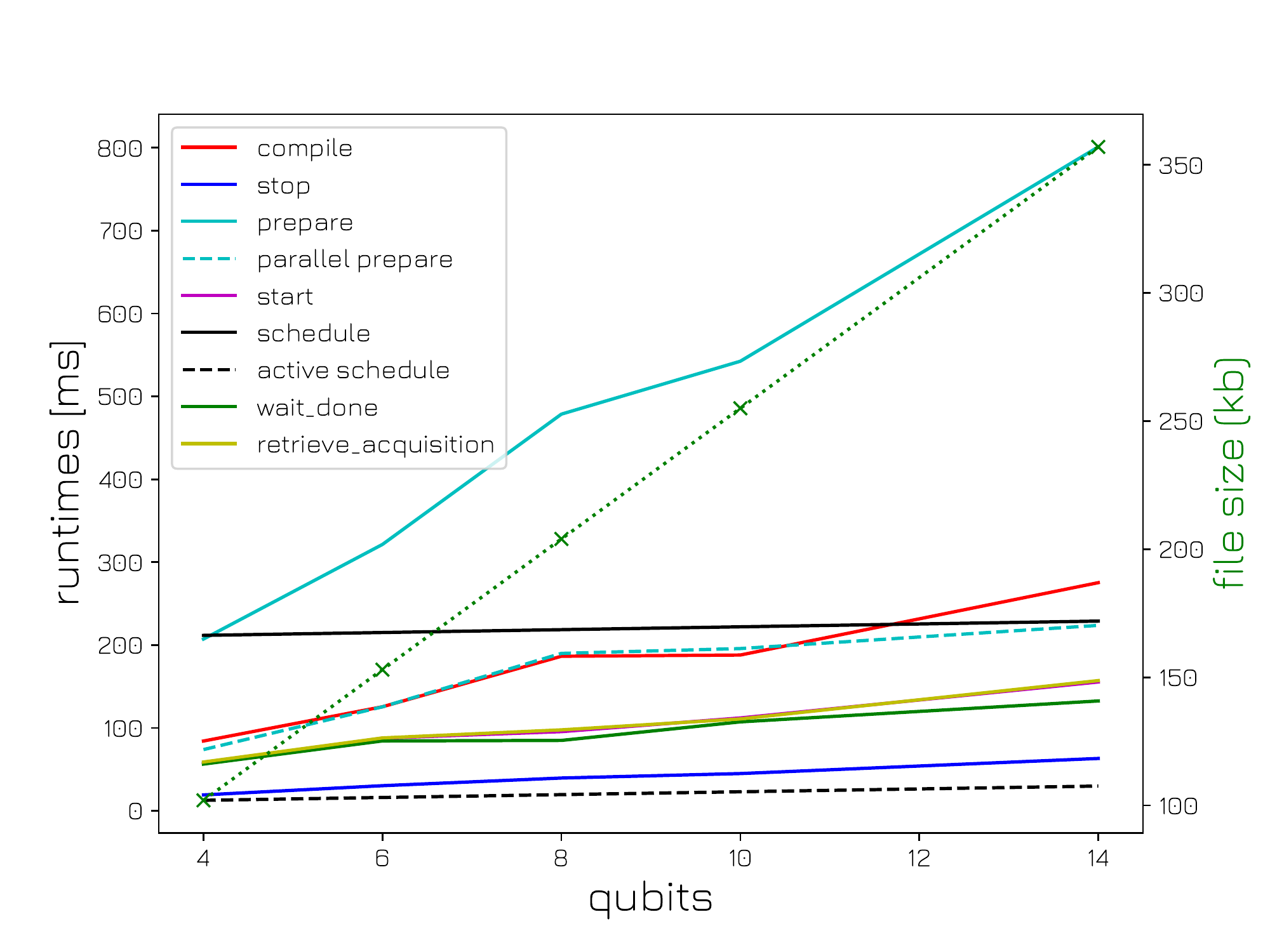}
    \caption{
    \textbf{Left axis}: Average runtimes per optimization iteration for QAOA on 4,6,8,10,14~qubits, each measured for a total of 1000~shots.
    Specifically, the results are averaged over the optimization iterations and over 40 runs of the benchmark.
    \textbf{Right axis}: schedule data size, see dotted line with cross markers.
    }
    \label{fig:results_14}
\end{figure}

We examine the growth of the runtimes as a function of the number of qubits in order to gain insight on bottlenecks for larger systems. 
The profiling results up to 14~qubits are presented in~\cref{fig:results_14}, showing that the hardware prepare step grows with the highest slope and is dominating the execution time. 
A speedup of 3.58x is achieved at 14~qubits when comparing the parallel and baseline prepare step.  
By implementing the parallel prepare, a linear growth of the execution time can still be observed, albeit small.
We attribute this linear growth to the growth of schedule data sent during the prepare command and to the time that it takes the modules to get ready.
Measurement of the data size shows that the data grows linearly with the number of qubits, at a rate of about 25~kb per qubit, correlating to the linear increase in prepare time.

\subsection{Extrapolation to 50+ qubits}
\label{sec:results_50Q}

As the scale at which variational quantum algorithms may be useful is around 400~qubits~\cite{guerreschi2019qaoa}, it would be interesting to extrapolate the current results to that scale.
However, we cannot simply make a linear fit of the measured data from 4 to 14~qubits and extending it that much since the order of growth (linear, quadratic or else) depends on many factors and assumptions.
We discuss our expectations for the QAOA~optimizer, schedule, prepare step, compilation and number of SWAPs in a square-grid connectivity.

As far as the optimizer is concerned, the number of iterations will highly depend on the smoothness of the cost function and the number of local minima.
These factors make it harder for an optimizer to find the global optimum, affecting the accuracy metric of the algorithm.
As this function is the black-box evaluation of the quantum algorithm, this will strongly depend on the problem instance and quantum-device noise.
However, two indications about the optimization time can be given.
For the SHGO~algorithm specifically, it is known that SGHO is very unlikely to give good results for over 10~variables~\cite{shgo}.
As the optimization of QAOA reaches this point already for small problem sizes, the performance of SHGO is likely not well suited for QAOA on large scale.
Current research explores the option of machine learning to solve this optimization issue, but results show that training these algorithms is very time-consuming, in the order of hours even for small problem sizes~\cite{khairy2019reinforcement,qaoa_ml}.
The second concern for the time complexity of QAOA optimization is the growth of parameters.
This is determined by (a) the number of constraints and (b) the number of QAOA layers.
Note that the total number of parameters is the multiplication of both factors (times a small constant).
The growth of layers as a function of problem size is up until now only determined empirically.
However, the number of constraints grows at worst quadratically ($\frac{n(n-1)}{2}$ where $n$ is the number of qubits) for a fully connected graph and at best linearly under the assumption that all nodes are connected. 

The potential quadratic scaling of the QAOA~parameters discussed above reflects directly in the schedule execution time.
The size of the schedule file would also grow accordingly, increasing the load time of the compiled schedules during the (parallel) prepare step. 
Up to 14~qubits we observe that the description of the model waveforms and the schedule itself, i.e.~the ordering and parameter specification of the waveforms to form the schedule, take different fractions of the data depending on the type of module.
For readout modules the data consists of $\sim 87.5\%$ waveform descriptions and $\sim 12.5\%$ schedule instructions (measurements).
These do not change for larger quantum circuits as there is only one final measurement per qubit.
For control modules the data consists of~$\sim 12.5\%$ waveform descriptions, which remain constant as no new types of gates are introduced in scaling the circuit.
The schedule description occupies~$\sim 87.5\%$ of the data, which can grow for larger systems, but hardly increases for this particular benchmark (we only observe that the per-qubit schedule description varies by at most 1~kb per instruction file)
We attribute this to the fact that, for the chosen problem instance in the used QPack benchmark~\cite{mesman2021qpack}, as discussed in~\cref{sec:qaoa}, this ratio of constraints to qubits is linear, therefore the per-qubit schedule does not increase on average (not counting SWAP~gate overhead, see below).
As the increase in file size due to increasing schedule length is negligible at this scale, the only driving factor is the linear increase of total data due to the linearly increasing number of compiled files (one control and one readout per qubit).
However, if the problem would be structured differently, the number of constraints per qubit could grow with the problem size.
As then only the control schedule itself may grow quadratically in the worst case,
we predict that the growth in data size and thus prepare-step runtime will still look linear for relatively low number of qubits, even though a specific number for the latter is hard to determine.
The possible factor that is expected to be most impactful on the prepare time is the loading of the schedules on the sequencers, which increases for larger problem sizes.
The bandwidth of the connection might create a bottleneck at some point, however, with file sizes in the order of 25~kb per qubit, the bandwidth of Ethernet connections (1~Gbps) will not cause concerns for near-term quantum computers (for example, it takes about~$360~\mu$s to transfer the schedules for 14~qubits, plus a small latency overhead). 

The scaling of the compilation is harder to predict.
The quadratic growth of the schedule will introduce a quadratic scaling of some part of the compilation step, but currently it is unclear how large the impact will be.
Furthermore, the given compile runtimes are for all-to-all connectivity and thus do not include the overhead due to the compilation of SWAP gates (see below) in a limited connectivity.
The latter is hard to isolate in Qiskit and to combine with the different schedule format in Quantify.

\begin{figure}
    \centering
    \includegraphics[width=\linewidth]{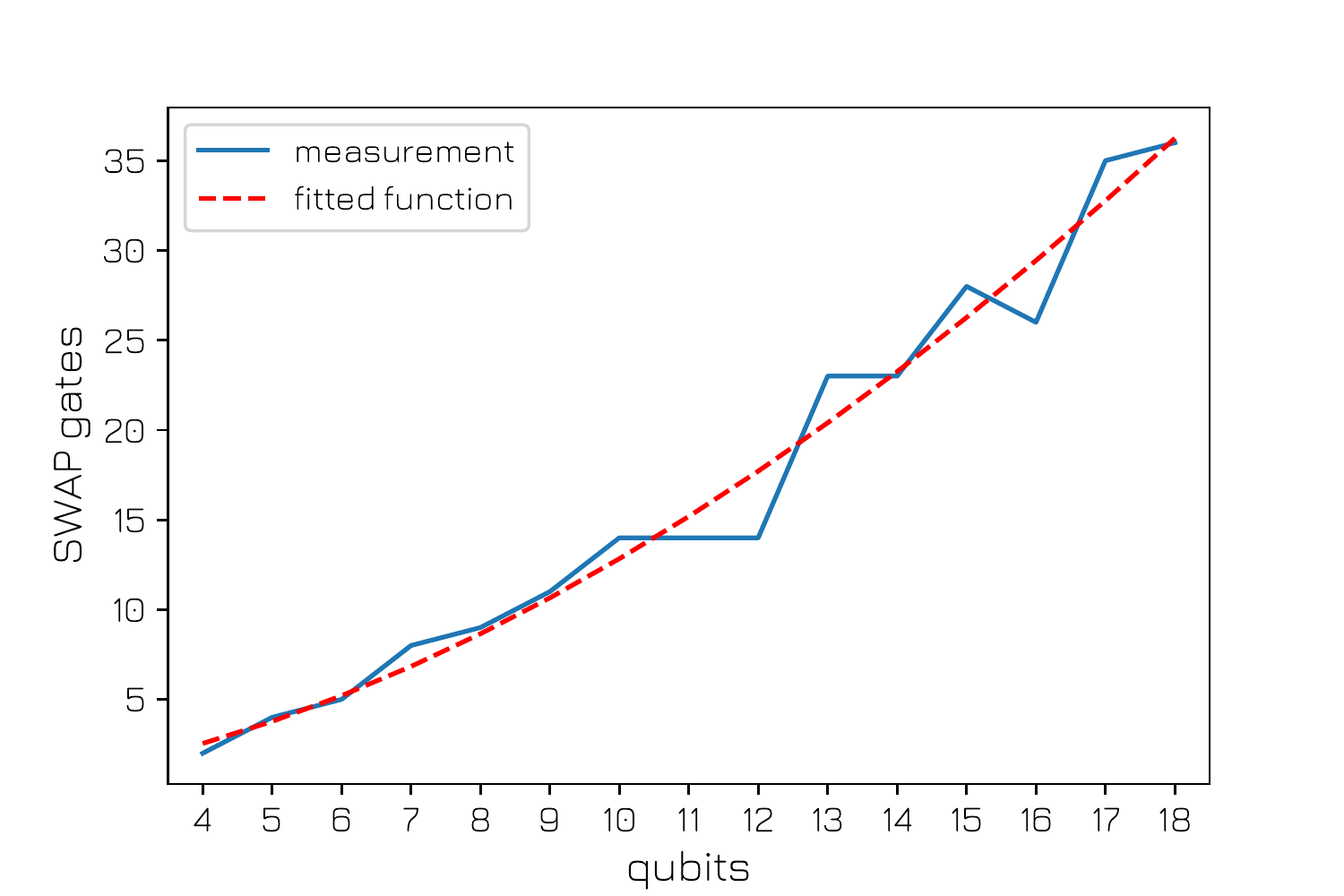}
    \caption{
    Measured results for the number of SWAP~gates required for the QAOA~algorithm on a square-grid qubit layout.
    The results are fitted to a polynomial function, obtaining an exponent of~1.7.}
    \label{fig:swaps}
\end{figure}

As we have considered all-to-all connectivity for simplicity but this is typically not available in transmon qubits, each two-qubit gate has to be compiled making the use of a number of SWAP~gates.
This is dependent on the problem instance, the number of constraints and the qubit layout, leading in general to an increase in the schedule time.
In order to estimate the growth of this overhead, we use Qiskit's schedule transpiler \cite{qiskit_transpiler} for the same QAOA~algorithm to compile the SWAP~gates for a square-grid layout. 
This layout has been chosen to emulate the chip in Ref.~\cite{Krinner21} as well as Google's Sycamore processor~\cite{Arute19}.
The results are shown in~\cref{fig:swaps}. 
We fit them to a polynomial function, resulting in a power of~1.7. 
The found growth of the number of SWAP~gates is realistic, as it is at most cubic (quadratic number of two-qubit gates times maximum linear number of SWAPs per two-qubit gate) and with a minimum of~0 (in the case the problem maps exactly to the qubit layout).

As discussed in this section, the extrapolation of the measured average runtimes is uncertain given the unknown factors in the scaling of the QAOA~layers, schedule size and compilation.
Nevertheless, we perform a best-case-scenario, linear extrapolation to 50~qubits to get a rough estimate of the required runtimes (see~\cref{tab:50q}).
Only the schedule time includes the polynomial extrapolation of the additional SWAP~gates. 
These numbers indicate that the next major bottleneck is compilation (in fact, this is the case in~\cref{fig:results_14} already for 14~qubits). 
As stated before, the actual time is expected to be even larger compared to the linear estimation.
In~\cref{sec:discussion} we propose ways to mitigate this compilation bottleneck.

\begin{table}
    \centering
    \caption{Linear extrapolation to 50~qubits of the measured average runtimes per optimization iteration, in descending order, for 1000~shots each, for the case with active reset and parallel module preparation.}
    \begin{tabular}{|c|c|}
    \hline
    \textbf{Execution step} & Runtime [s]\\
    \hline
    \hline
    Compile & 0.923 \\
    Parallel prepare & 0.754 \\
    Start & 0.479 \\
    Retrieve acquisition & 0.478 \\
    Wait done  &  0.388  \\
    Stop & 0.213 \\
    Schedule (active reset) & 0.090 \\
    \hline
    \hline
    Total  &  3.325  \\
    \hline
    \end{tabular}
    \label{tab:50q}
\end{table}

\section{Discussion} \label{sec:discussion}
While the presented runtime measurements break the runtime into its main categories, Q-Profile is capable of further decomposing them into finer blocks. 
Further profiling the computation steps is a necessity as new bottlenecks will emerge.
One of the expected bottlenecks is the quantum-circuit compilation, as discussed in~\cref{sec:results_50Q}.
This step currently occurs from scratch for every optimization iteration.
However, the quantum circuits contain the same gates with different parameters only. 
One solution would be to parameterize the quantum-circuit schedule, where only the updated parameters are uploaded to the control hardware while the general circuit structure remains stored on the control hardware. 
This would significantly reduce the quantum-circuit compilation time and further accelerate variational quantum algorithms.
Optionally, dedicated hardware acceleration can be considered.
Note that if one would optimize the schedule to use parallel gates to best fit the qubit layout, more time for compilation would be required than here. 
Quantum circuit optimization is a challenging task in itself, with contextual circuit improvements for e.g.~QAOA~\cite{qaoa_compile}. 
Including this in the compilation would give rise to even larger compile times, leading to a tradeoff between compile time and circuit optimization.

Our tool shares common aims with the CLOPS~benchmark~\cite{CLOPS}.
This benchmark aims at measuring the speed at which not only gates are executed on a quantum chip, but also the speed at which classical computation proceeds to schedule and compile instructions, as well as to collect results and perform post-processing.
IBM's results~\cite{CLOPS} indicate a bottleneck in the quantum control hardware which increases with larger devices, as more control hardware is required and as the initialization and load times increase. 
In particular, the IBM~cloud setup introduces most of its overhead in the communication, as could be expected, even to a much higher degree compared to our results.
As algorithm, the CLOPS benchmark utilizes parameterized quantum-volume (random) circuits in order to represent a typical use of the quantum computer.
Its score is influenced by the number of shots, templates, parameter updates and quantum-volume layers.
While our tool could be used to run the same circuits, in this work we chose instead to apply it to QAOA because we believe that profiling specific algorithms of broad interest is relevant and timely as well.

\section{Conclusion} \label{sec:conclusion}
We have proposed and demonstrated Q-Profile, an accurate and efficient tool for benchmarking and profiling quantum control hardware and software. 
Profiling quantum systems is becoming increasingly important due to the aim of using quantum computing for acceleration. 
We applied Q-Profile to QAOA to reflect near-term practical usage, but it can be extended to other quantum algorithms, as well as to more qubits to study how the overhead and its profile scale with system size.
The presented tool indicated various bottlenecks for the current state of (variational) quantum computing, and demonstrates capabilities for profiling quantum control stacks in the future. 
An improvement to the hardware control software has been implemented to demonstrate the effectiveness of profiling. 
With this work, we have demonstrated the utility of Q-Profile and given insight to hardware and software developers in how to find performance bottlenecks in their systems.
The profiling capability has been included in the open-source software Quantify~\cite{quantify}.
\bibliographystyle{IEEEtran}
\bibliography{bib}
\end{document}